\documentclass[aps,prb,twocolumn,groupedaddress]{revtex4}
\usepackage{graphicx}% Include figure files

\begin{document}

%\preprint{}

\title{On the heterogeneous character of water's amorphous polymorphism}

\author{M.M.~Koza,
        R.P.~May,
        H.~Schober.}

\affiliation{Institut Laue Langevin, 6 Rue Jules Horowitz, B.P. 156, 
38042 GRENOBLE, Cedex 9, France.}

\begin{abstract}
In this letter we report {\it in situ} small--angle neutron scattering
results on the high--density (HDA) and low-density amorphous (LDA) ice
structures and on intermediate structures as found during the temperature
induced transformation of HDA into LDA.
We show that the small--angle signal is characterised by two $Q$ regimes
featuring different properties
($Q$ is the modulus of the scattering vector defined as
$Q = 4\pi\sin{(\Theta)}/\lambda_{\rm i}$
with $\Theta$ being half the scattering angle and $\lambda_{\rm i}$
the incident neutron wavelength).
The very low--$Q$ regime ($<~5\times 10^{-2}$~\AA $^{-1}$) is dominated by a
Porod--limit scattering.
Its intensity reduces in the course of the HDA to LDA transformation
following a kinetics reminiscent of that observed in wide--angle
diffraction experiments.
The small--angle neutron scattering formfactor in the intermediate
regime of $5 \times 10^{-2} < Q < 0.5$~\AA$^{-1}$
HDA and LDA features a rather flat plateau.
However, the HDA signal shows an ascending intensity towards
smaller $Q$ marking this amorphous structure as heterogeneous.
When following the HDA to LDA transition the formfactor
shows a pronounced transient excess in intensity marking all
intermediate structures as strongly heterogeneous on a length scale
of some nano--meters.\\

{\bf Final and complete version is published
in: J. Appl. Cryst. 40, s517--s521, (2007)
doi:10.1107/S0021889807004992}

\end{abstract}

\maketitle

Throughout recent years different concepts have been introduced to explain
the phenomenon of amorphous polymorphism, i.e. the existence of more
than one amorphous structure in a single substance
(Mishima, 1998; Stanley, 2000; Debenedetti, 2003).
The most intriguing scenario is based on the existence of two distinct
liquid states, that has been established as a possibility
in molecular dynamics simulations (Poole, 1993).
The liquid polymorphism was supposed to account for the formation of the
high--density amorphous (HDA, $\rho \approx 39$~molecules/nm$^3$)
and the low--density amorphous (LDA, $\rho \approx 31$~molecules/nm$^3$)
ice structures as the quenched liquid phases.
This scenario has been successfully extended towards other systems,
indicating that amorphous polymorphism, as a manifestation of distinct
liquid states, could be a general feature of condensed matter
(Kurita, 2004, 2005).
However, it has been equally questioned by recent computer
experiments that introduced bandwidth of transformation scenarios
spanning between the extrema of a pure relaxation phenomenon
of an amorphous matrix and a multiple--phase transition scheme
(Guillot, 2003; Brovchenko, 2003; Martonak, 2004).

The experimental proof or counterevidence of the two--liquid scenario
in water is a subtle task, as any attempt to directly access
the hypothetical two--liquid regime is bound to fail.
Fingerprints of liquid polymorphism, thus, are looked for in
the amorphous states.
One crucial indicator is the presence of a first--order transition
between HDA and LDA.
However, any experimental approach towards a classification of the
HDA--to--LDA transformation is severely hampered by the non--ergodic
nature of the amorphous structures (Koza, 2005a).

Hence, an experimental search for characteristic
features that might help to discern between the proposed thermodynamic
concepts is the only approach to shed some light on the origin of
the amorphous polymorphism of ice.
One characteristic feature of the HDA--to--LDA transformation
is an enhancement of the elastic signal in the small--angle
scattering regime.
Already the very first {\it in situ} studies of the HDA--to--LDA
transformation have shown that despite the continuously changing static
and dynamic structure factors of the amorphous ice there is also a
transient excess signal at low scattering angles
(Schober, 1998; Schober, 2000).
The enhanced small--angle neutron scattering (SANS) and 
small--angle X--ray scattering (SAXS) signals could be understood as
the response of frozen--in density variations, i.e. spatial heterogeneities,
occurring transiently during the structural transformation of
HDA into LDA.

In this manuscript we report on {\it in situ}
neutron diffraction experiments in the small--angle scattering
regime monitoring structures of typically 10--1000~\AA .
From a set of time and temperature dependent experiments,
we are able to substantiate the transient, thoroughly heterogeneous
character of the intermediate transition stages on a spatial scale
of some nano--meter and to describe qualitatively the temperature
dependence and time evolution of the transient heterogeneities.
We will equally show, that beyond the $Q$ regime of the transient
enhanced scattering there is a signal due to interfaces
whose kinetics follows the wide--angle response (Koza, 2003).

All samples were prepared by slow compression of D$_2$O (purity 99.8~\%)
ice~I$_{\rm h}$
at $T\approx 77$~K up to pressures of 1.8~GPa in a piston cylinder
apparatus (Koza, 2003).
Each preparation run resulted in a sample volume of 3~ml.
To meet the condition of a 10~\% scattering sample for the suppression
of multiple scattering, the samples were carefully crushed into milli--meter
size chunks and placed within a cadmium diaphragm into a standard flat
aluminium sample holder.
The diaphragm optimised the size of the sample, leaving a free space
of 12~mm in diameter and about 2.2~mm thickness, at its centre with respect
to the homogeneous neutron beam.

The purity of the HDA samples had been confirmed at the spectrometer IN6
($0.3$~\AA $^{-1} \le Q \le 2.7$~\AA $^{-1}$)
before they were mounted for measurements in a standard
cryostat at the small--angle diffractometer D22.
$Q$ is the modulus of the scattering vector defined as
$Q = 4\pi\sin{(\Theta)}/\lambda_{\rm i}$ with $\Theta$ and
$\lambda_{\rm i}$ being half the scattering angel and
the wavelength of the incident radiation, respectively.
Both instruments are situated at the Institut Laue Langevin
in Grenoble, France.
The transmissions of the samples were determined experimentally
in the HDA state at D22 to $89\pm 2$\% corresponding with an
effective sample thickness of $1.5\pm 0.3$~mm.
Samples had been kept for about 30~min at 78~K for a complete
removal of liquid Nitrogen, before any high accuracy measurement
was started.
During the experiments an atmosphere of 200~mbar of helium
was applied.

The measurements were carried out with an incident
neutron wavelength of 6~\AA\ and additional test measurements
were carried out with a wavelength of 24~\AA\ in order to access
the largest structural units.
A lowest limit of $Q \approx 5\times 10^{-4}$\AA $^{-1}$ could be
reached.
The exploitation of such a large $Q$ range
requires in practice the modification of the instrumental setup.
High accuracy measurements in the structurally stable states
HDA and LDA were carried out with three detector to sample distances,
namely 1.4~m, 5~m and 18~m.
Data acquisition time was 10~min at each position.
The setup choice for {\it in situ} studies of the transforming
structures was contingent upon the transformation kinetics at
the chosen temperatures since a dead time of 1.72~min. was
due to the setup changes.
Consecutive measurements were performed with a detector to sample
distance of 1.4~m and 10~m for 3~min. each.
At the temperature 105~K only a single detector to sample
distance of 2.5~m could be applied due to the
fast transformation kinetics of the sample.

In this paper we can only present a subset of our experimental data.
The four samples reported here were followed {\it in situ}
at the nominal temperatures of 100~K, 101.5~K,
103~K and 105~K.
Please note that throughout this paper we refer to the sample states
measured prior to the heat treatment as HDA and after having followed
the transformation and an annealing procedure at 127~K as LDA.
No changes of the LDA structure factor could  be  found upon
cooling of the samples back to $\approx$~78~K.

Standard data corrections for empty can and background scattering were applied
for the setups with 1.4, 2.5, 5 and 10~m.
The calibration of the detector and normalisation to absolute
units were accomplished with a water (H$_2$O) standard of 0.1~mm
thickness.
The effective scattering power of the water standard at $\lambda = 6$~\AA\
was taken into account (Lindner, 2002).
All corrections and the azimuthal averaging of the two--dimensional
data were
done with the software package GRASP (Dewhurst, 2003).
For a clear presentation the data sets were
normalised to unity with respect to an LDA baseline.
The normalisation factor onto an absolute scale is
$4.8(1)\times 10^{-2}$~cm$^{-1}$.

\begin{figure}
\caption{\label{fig_jac_00}
Diffraction data recorded in the course of the HDA to LDA
transformation.
The small--$Q$ intensity indicates a transient intensity excess
and a state of strongest heterogeneity (SSH) can be identified.}
\includegraphics[angle=0,width=85mm]{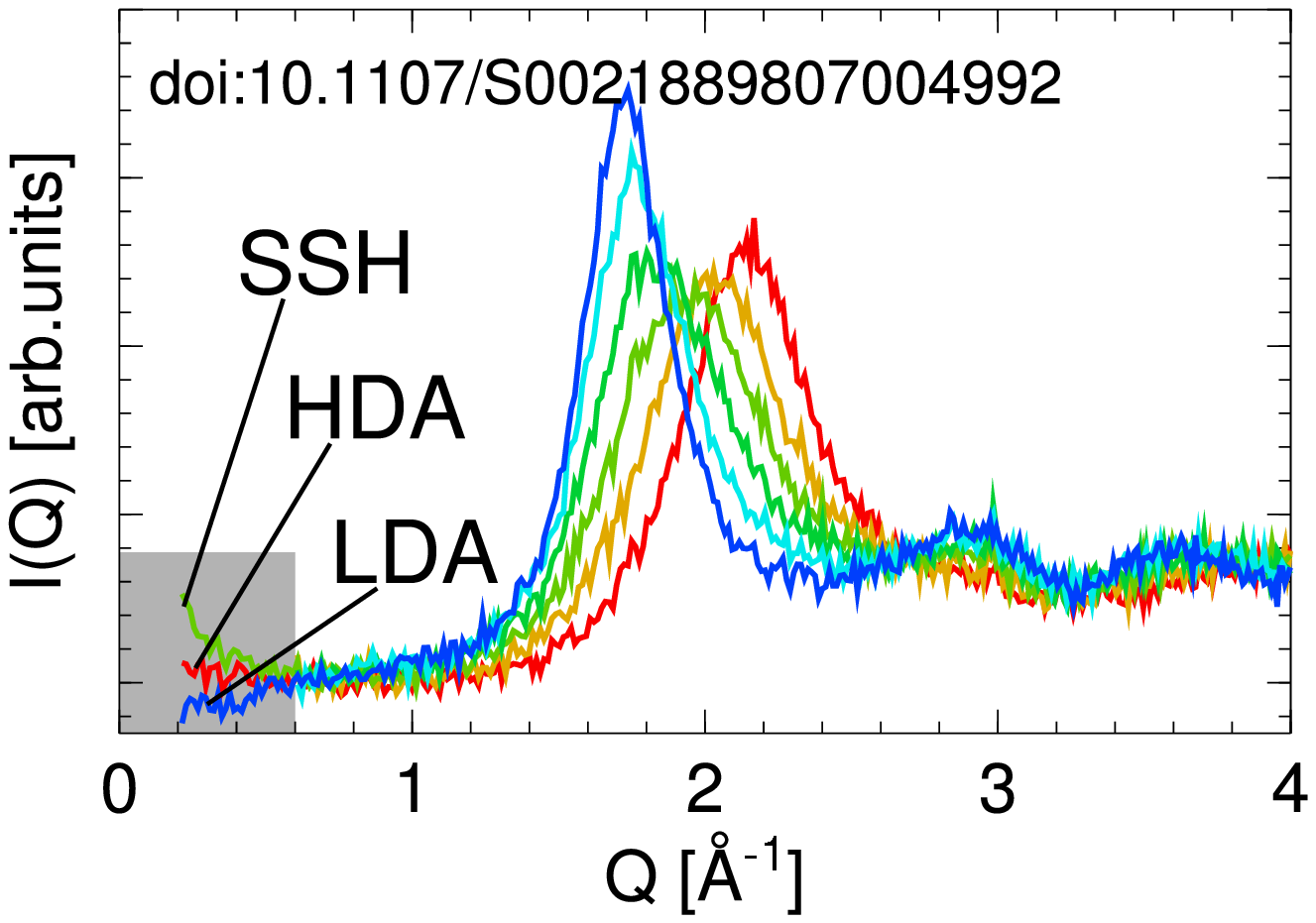}
\end{figure}

For the readers' convenience we report in figure~\ref{fig_jac_00}
a selected set of wide--angle diffraction data recorded in the course
of the HDA to LDA transformation at the diffractometer D20 at Institut
Laue Langevin.
The grey shaded area stresses the small--$Q$ regime in which
intensity changes indicate that the sample passes through
a state of strongest heterogeneity (SSH). 
Please note that already the HDA structure displays an
intensity higher than the one of LDA which is in agreement
with prior publications for neutron and X--ray scattering
(Schober, 1998; Schober, 2000; Koza, 2005b; Koza, 2006).
The small--$Q$ signal of the other intermediate tranformation
stages has been suppressed for a clear presentation.

\begin{figure}
\caption{\label{fig_jac_01}
Left: SANS intensity $I(Q)$ of four different samples in the HDA
and LDA sample state.
Data have been shifted for clarity.
Solid lines represent fits to the data with a PLS and a constant 
background fixed to unity as given in the experimental section. 
Right: A close up of the $I(Q)$ in the intermediate $Q$ range.
Equally shown is the signal recorded on IN6 and reported
in (Schober, 1998).
The data have been shifted for clarity.}
\includegraphics[angle=0,width=85mm]{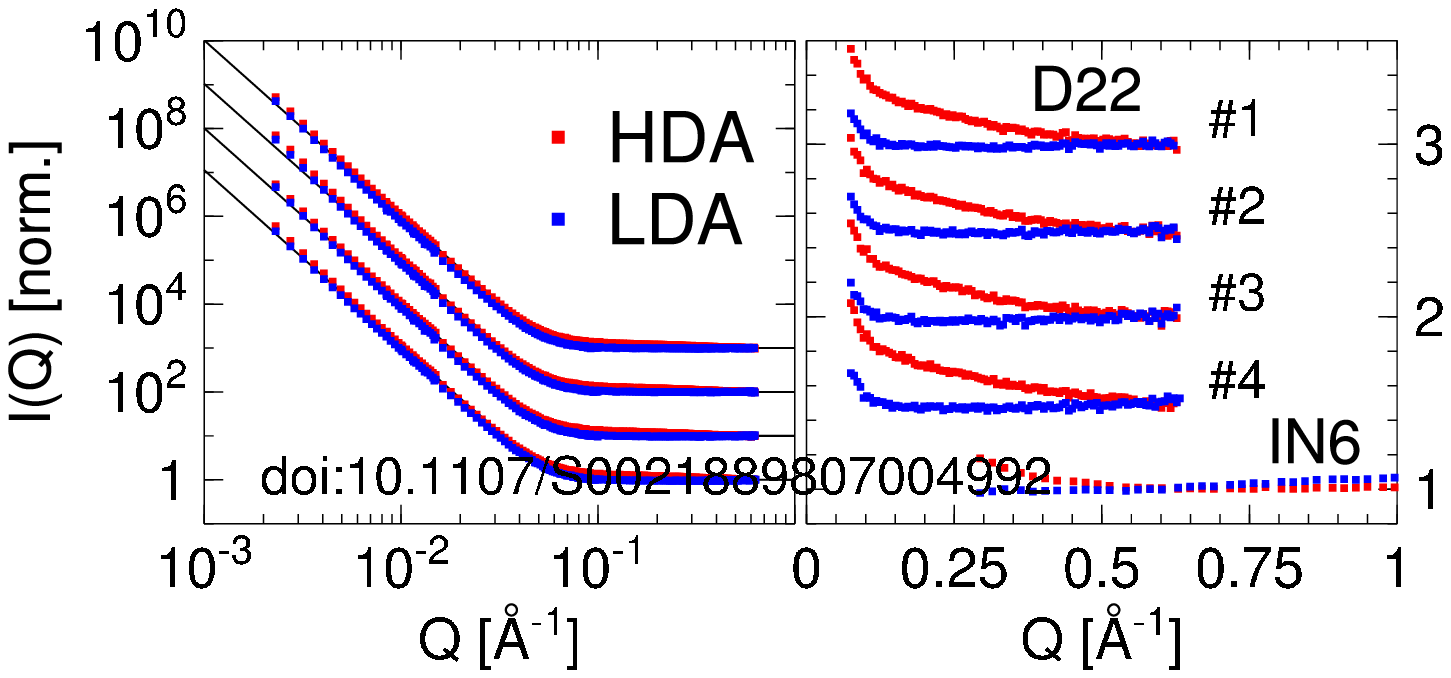}
\end{figure}

Small--angle scattering data taken at D22 are shown in 
figure~\ref{fig_jac_01}.
In the left panel we report the
intensities of all samples measured in the HDA and the LDA states.
Data sets of different samples are shifted for clarity.
Two features dominate the signal, an apparently flat background
at $Q > 5\times 10^{-2}$\AA $^{-1}$, comprising contributions from
the incoherent scattering from D$_2$O and from density variations
that can be associated with the compressibility of the amorphous
matrix, and a power law scattering towards smaller $Q$.
The pronounced power law dependence $S(Q)\propto Q^{\rm -4}$ is the
so called Porod--limit scattering (PLS).
It is the final slope of a SANS form factor
that appears due to a sharp boundary
between two phases in a sample and depends only on the scattering  
contrast and the interface area, but not on the shape of the  
structures or particles present in the sample (Glatter, 1982; Lindner, 2002).
Note that the intensity of the PLS was well reproduced
in all our samples following the same preparational procedure.
At low Q, our data do not cover the range necessary to observe a
Guinier-limit scattering.
Furthermore, comparison of results obtained with 6 and 24~\AA\ show
a clear influence of multiple scattering on the data 
(Lindner, 2002).
The intensities of the PLS were determined to 
$4.8(1)\times 10^{-15}$~\AA $^{-5}$ and $3.7(1)\times 10^{-15}$~\AA $^{-5}$
for HDA and LDA, respectively.
Please see the Appendix for more information.

Let us focus in the following on the momentum range $Q > 0.1$\AA~$^{-1}$,
i.e., at the apparently flat background beyond the PLS.
Right panel of figure~\ref{fig_jac_01} reports the intensities
of the four samples in comparison to prior results obtained on
the spectrometer IN6 (Schober, 1998).
As it has been shown before in X--ray and neutron scattering
experiments (Schober, 1998; Schober, 2000; Koza, 2005b; Koza, 2006)
an excess of the SAS signal indicates a pronounced heterogeneous
character of the initial HDA structure.
Whereas the LDA modification shows a constant signal.
These features are entirely reproducible.

\begin{figure}
\caption{\label{fig_jac_02}
Left: Intensity evolution $I(Q,t,T=103$~K) measured at different
stages of the HDA to LDA transformation with sample \#2.
As it is indicated by the vertical arrows the top figure reports
the increase of the transient signal, the bottom figure shows its downturn.
Time $t$ after which the signal has been recorded is indicated
in the figure.
Solid lines are to be taken as guides to the eye. 
Right: Corresponding signal evolution measured with sample \#1 at 105~K.}
\includegraphics[angle=0,width=85mm]{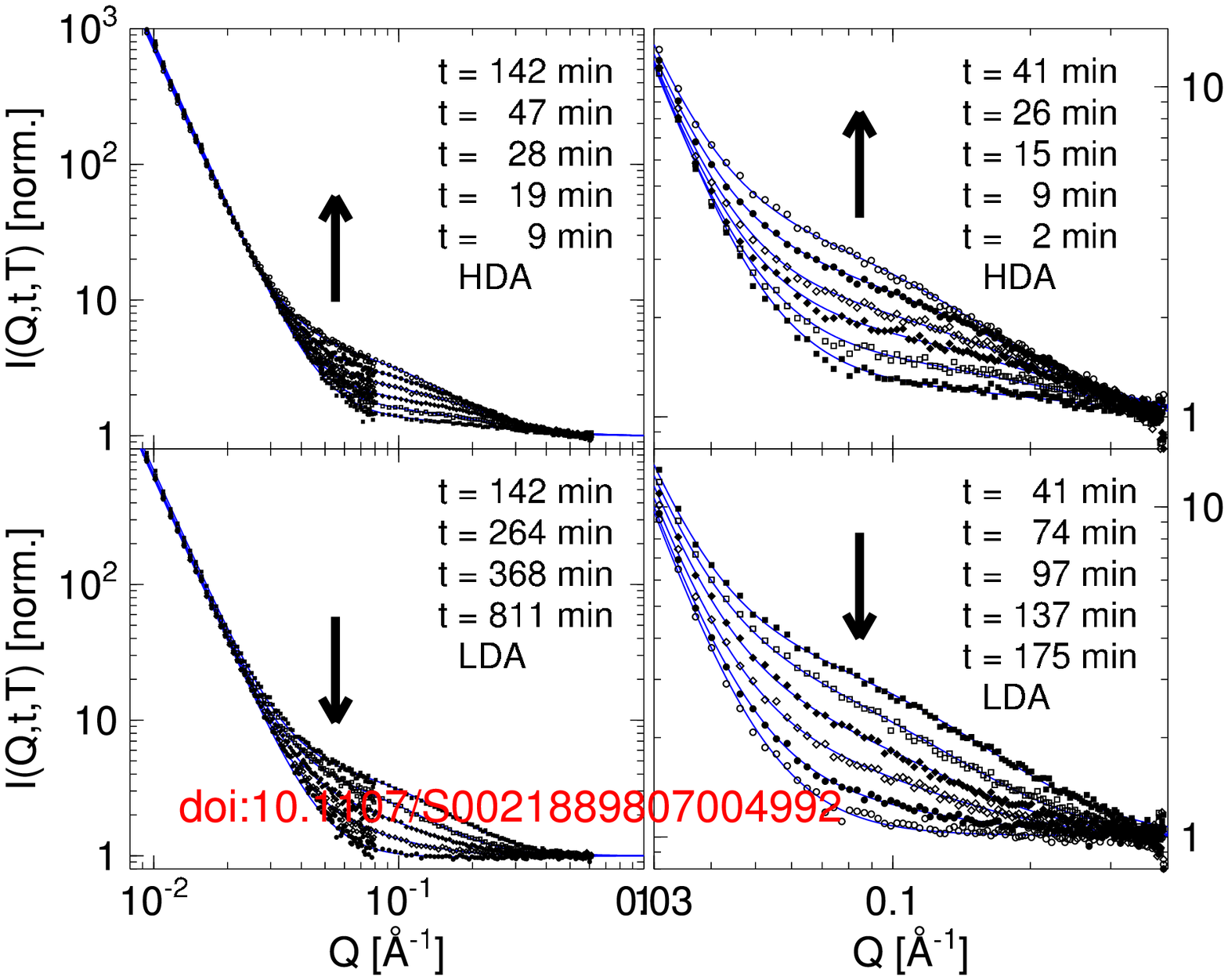}
\end{figure}

Figure~\ref{fig_jac_02} reports the {\it in situ} SANS formfactors $I(Q,t,T)$ 
of samples \#2 (left hand side) and \#1 (right hand side) in a double--log
plot.
As it is indicated by the vertical arrows the increase of the
transient signal is plotted in the top figures, its downturn is shown at
the bottom.
Equally indicated is the time $t$ after which the data
have been recorded.

It is evident that $I(Q,t,T)$, in the presented SANS regime,
is a characteristic measure of the HDA--to--LDA transition,
In analogy to the features of the wide--angle diffraction (WAD) signal
elaborated in reference (Koza, 2005a), we can find
for example to any $I(Q,t,T=103\rm K)$ a matching signal observed
at a different $T$ after a well-defined but different evolution time $t$.
As a consequence, it is not only the mere presence of a SANS
signal that is characteristic of the intermediate structures,
but it is in particular its intensity and the details of
its profile that discern and, basically, define distinct
transition stages.

\begin{figure}
\caption{\label{fig_jac_03}
Left: Kinetic response of sample \#2 in the Porod--limit scattering
$I_{\rm P}(t,T=103$~K) and the intermediate $Q$ regime $I_{\rm I}(t,T=103$~K).
Right: Corresponding response of sample \#1 at 105~K.
Please note that the fall off the signals at the end of data sets is due to
the annealing process of the samples to LDA at ~127~K.}
\includegraphics[angle=0,width=85mm]{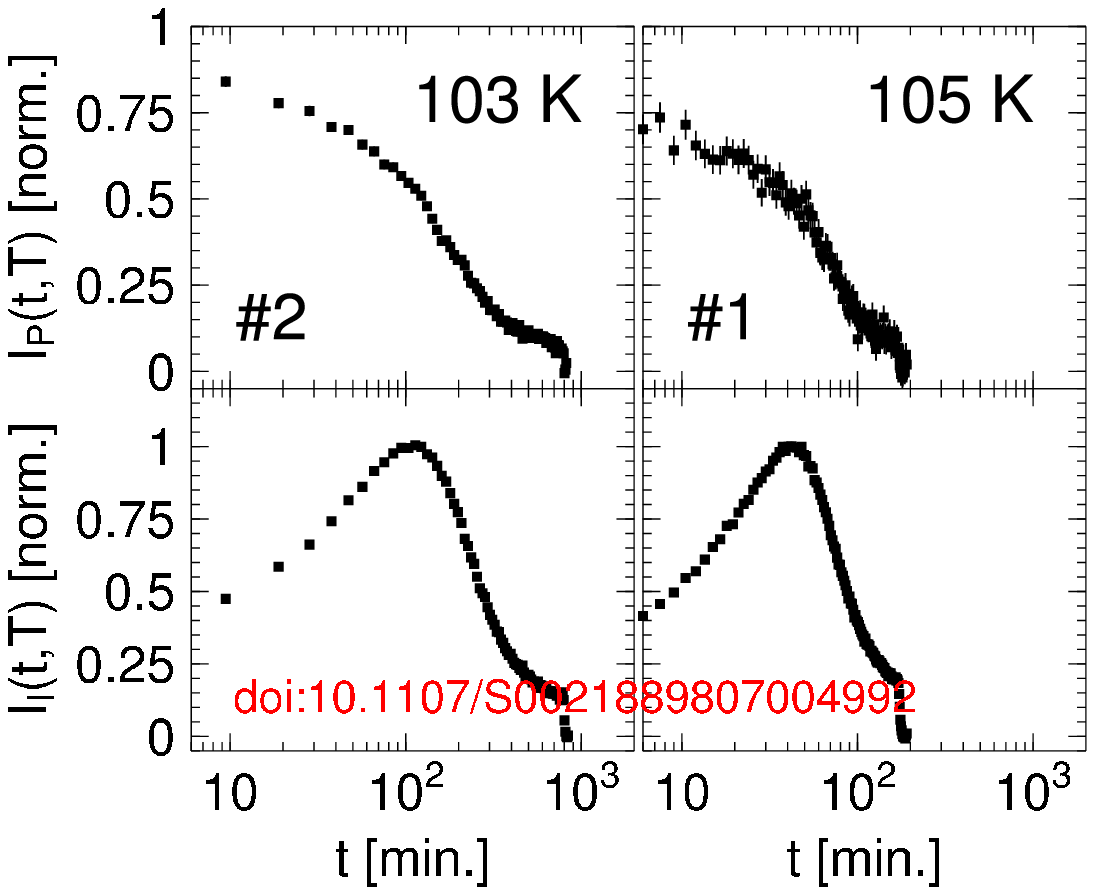}
\end{figure}

A similarity with the wide--angle signal can be also found
in the kinetic properties computed from $I(Q,t,T)$.
For simplicity we restrict our consideration here to intensities
integrated over two different $Q$--ranges.
One range ($10^{-2}\le Q \le 2\times 10^{-2}$\AA $^{-1}$) stresses
the PLS evolution ($I_{\rm P}(t,T)$) and the other
($0.1\le Q \le 0.15$\AA $^{-1}$) represents the $Q$ regime
of the transient excess scattering ($I_{\rm I}(t,T)$).
Figure~\ref{fig_jac_03} shows the time dependence of $I_{\rm P}(t,T)$
(top panel) and $I_{\rm I}(t,T)$ (bottom panel) of the two samples.
The data have been normalised in accordance to reference
(Koza, 2005b).
$I_{\rm P}(t,T)$ takes on values between unity, representing the HDA state,
and null, representing the LDA state.
$I_{\rm I}(t,T)$ is defined as null for the LDA state and unity
for the SSH.
Note that the relaxed statistics of the $I_{\rm P}(t,T=105$~K)
signal is due to the limited $Q$--range given by the 
detector to sample distance of 2.5~m.

$I_{\rm P}(t,T)$ displays a dependence well comparable with the evolution
of the WAD signal reported in detail in reference~(Koza, 2003).
Two features are dominating the time response.
First, a sluggish transformation process is observable, and second,
a sigmoid shaped (Avrami--Kolmogorov type) step is detectable.
Although, $I_{\rm P}(t,T)$ bears fingerprints of an Avrami--Kolmogorov
type transformation the entire process cannot be understood
as a simple nucleation and growth scenario (Doremus, 1985).
This has been discussed in detail in ref.~(Koza, 2003).

We may find a simple explanation for the equivalent behaviour 
of the PLS and WAD kinetics.
Taking into account that the intensity of the PLS is proportional
to the square of the difference in scattering density and the
specific surface $\Delta\rho^2\sim S/V$, it is an index
of the density of the sample.
In an equivalent way a dependence has been established between the
density of amorphous
samples and the relative position of the structure factor maximum, however,
determined by structural changes on a local length scale of some \AA\
(Elliott, 1991, 1995). 

It has been indicated recently based on diffraction experiments
(Koza, 2005b) that the kinetics of the SANS signal in the intermediate
$Q$ range is equally closely related with the WAD.
Here we show that this relation applies as well to the PLS kinetics.
The SSH can be found always close to the centre of the transformation
between HDA and LDA.
This behaviour is independent of $T$.
However, one has to take some care when interpreting $I_{\rm I}(t,T)$
in detail.
The decisive and accurate observable constitutes the Porod--invariant
$q=\int_0^{\infty}I_{\rm h}(Q) Q^2 {\rm d}Q$
of the formfactor $I_{\rm h}(Q)$ which characterises
exclusively the transient excess scattering.
This formfactor is dwarfed by the PLS intensity towards
smaller $Q$.
Nevertheless, a relation between the SSH position and
the centre of the transformation has been unequivocally
established in all our experiments accessing the intermediate
$Q$ range and either the PLS or the WAD signals.

The present SANS data establish unequivocally the thoroughly
heterogeneous nature of the amorphous ice structures when
following the transformation from HDA to LDA.
This characteristics exclude the possibility of a homogeneous
relaxation process of an amorphous ice matrix.
Hence, to explain the transformation behaviour we may think
of two other simple scenarios.

First, taking the non--ergodicity of the amorphous ice structures
into consideration we may think of the sample as being composed
of sub-ensembles each of which is governed by distinctly
different relaxation dynamics, i.e.\ relaxing for a given $T$
on different time scales.
This heterogeneity in relaxation behavior translates into a strong
 spatial heterogeneity of the system while going through the transition,
the reason being the large density differences between the still present
high-density and already relaxed low-density sub-ensembles. 

Second, a first--order transition may not be excluded as
a process underlying the HDA to LDA transformation.
Since this transformation is accomponied by an appreciable
density change of almost 30~\% the kinetics of the transformation
is expected to be strongly perturbed by the additional elastic
energy contribution as discussed in reference (Tanaka, 2000).
In particular we may expect that early transition stages encountering
molecules within a low--density environment surrounded by a high--density
matrix will be strongly stressed.
On a local scale, the sample is influenced by a non uniform
pressure distribution leading to departures from the
properties of a non--stressed bulk low--density amorphous structure.

Irrespective of the scenario, i.e. a heterogeneous relaxation or
a real phase transition, underlying the transformation between
a high--density and a low--density amorphous structures it is
obvious that HDA as it is prepared by compression at 77~K
is a heterogeneous structure on a nano--meter scale.
Hence, it is tempting to consider the very--high density amorphous
ice modification as the initial stage of the transformation
(Mishima, 1996; Loerting, 2001; Koza, 2005b; Koza, 2006).
As a consequence, the heterogeneous character of HDA
has to be properly accounted for when structural properties
are computed or modelled in real space from experimental data.
The pronounced small--angle signal should in general be a help
in descerning between different models trying to explain
the phenomenon of amorphous polymorphism.

The overall behaviour reported here on amorphous ice modifications
is not unlike the properties reported on a different system showing
apparently amorphous polymorphism, namely triphenylphosphite (TPP).
A thorough heterogeneous character of the TPP sample passing through
a phase transition between two homogeneous states has been established
by nuclear magnetic resonance, light--scattering and SANS experiments
(Senker, 2005).
TPP SANS data show a pronounced PLS and an excess signal at
intermediate $Q$ in the amorphous state (Alba--Simionesco, 2000).
Light scattering data confirm the transient heterogeneous
nature of the sample on a micro--meter scale and 
indicate a complex kinetics of the transition
which can deviate from an Avrami--Kolmogorov nucleation and growth
scenario when the transition happens via a spinodal decomposition
(Kurita, 2004).
Moreover, properties of intermediate stages cannot be
reproduced by a superposition of properties of the initial and
the final transition states, i.e. the superposition principle fails.

The features established during the spinodal decomposition
in TPP signify the complexity of a transition between amorphous
structures, which might be equally the case for amorphous ice.
This findings force us to conclude that superposition principles,
isosbestic point criteria or classical nucleation and growth
scenarios are of no particular significance when trying to
account  for the real origin and nature
of the transformation between amorphous ice structures. 

We have applied small--angle neutron scattering (SANS) techniques
to study the structural properties of amorphous ice modifications
on mesoscopic lengthscales.
It has been shown that the high--density amorphous (HDA) ice
produced by compressing crystalline ice is a heterogeneous structure
on a spatial scale of some nano--meters.
When following the transformation of HDA into the low--density
amorphous modification (LDA) the SANS signal displays a contrast
maximum at about the center of the transformation.
Thus, the sample passes through a state of strongest heterogeneity.

As it has been reported earlier and shown here in detail the transient
SANS signal is a characteristic feature of the HDA to LDA transformation,
and it is intrinsic to structures intermediate with respect
to the very--high--density amorphous (vHDA) modification and LDA.

When following the HDA--to--LDA transformation {\it in situ}
the evolution of the Porod--limit scattering shows a time dependence
reminiscent of the WAD signal (Koza, 2003).
Its kinetics cannot be described by a pure Avrami--Kolmogorov
time dependence, that characterises a plain nucleation and growth
scenario.
We have pointed out and discussed in detail that the non--applicability
of a nucleation and growth scenario does not exclude a real phase
transition of first order between two amorphous ice structures.
We may only draw the conclusion that a homogeneous relaxation
of an amorphous matrix is to be excluded as a possible transformation
scenario between high--density and low--density amorphous ice.  

An obviously important question is the origin of the strong PLS
in the samples.
The PLS persists on the explored $Q$ and time scales of the experiments
not only beyond their recrystallisation to ice I$_{\rm c}$
(Koza, 2005a; SANS data not shown here)
but also upon annealing HDA into the very-high density modification
(Koza, 2006).
We have undertaken efforts to reduce the PLS intensity, e.g. by
different sample treatments.
For example we have measured HDA disk samples of about 1~mm
thickness and 12~mm diameter before and after crushing them
into mm-sized chunks.
The effect of the sample consistency did not effect the PLS
intensity appreciably.
If we consider a scenario of uniform, spherically shaped
heterogeneous domains as the source of the PLS
and approximate the scattering densities by the
sample--to--vacuum contrast we may estimate the size of the
domains to 1--10~$\mu$m.
Hence, they are well separated by at least two orders of magnitude
from the transient structural changes on the intermediate 
scale and sufficiently large to accomodate crystallites
of ice I$_{\rm c}$ after a recrystallisation of LDA.

\begin{figure}
\caption{\label{fig_jac_PLS}
Contrast plot of the two--dimansional signal measured with three
samples having been precompressed to 0~GPa (a), 0.9~GPa (b) and 1.8~GPa (c),
respectively.}
\includegraphics[angle=0,width=80mm]{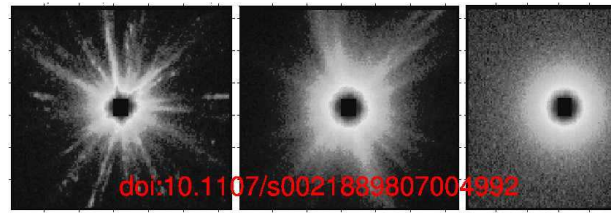}
\end{figure}

The consistent reproducibility of the PLS indicates that it
is either a generic feature of the amorphous ice samples or is created
by the compression process of the crystalline ice matrix.
For this reason we have examined crystalline samples that
had been precompressed to different pressures.
Figure~\ref{fig_jac_PLS} reports example data from three different runs.
The first crystalline sample has been formed within the pressure
device as for the preparation of the amorphous structure
at 77~K, however, no pressure was applied.
The second sample has been precompressed to 0.9~GPa
which corresponds to a pressure close to the formation
of HDA.
Figure \ref{fig_jac_PLS}~c reports the signal measured with one
of the HDA samples having been compressed to 1.8~GPa.

All our test runs showed a pronounced presence of impurities, dislocations
and stacking faults already within the uncompressed crystalline
samples.
This is visualized in figure \ref{fig_jac_PLS}~a by the anisotropic
scattering characteristic.
By applying pressure to the samples the signal from the perturbed
crystalline matrices indicated a trend towards isotropic scattering
(figure \ref{fig_jac_PLS}~b)
whereby a complete isotropic characteristics was reached in the
HDA structures (figure \ref{fig_jac_PLS}~c).
Although, this observation is based on {\it ex situ} compression
runs it indicates that the PLS is a generic feature of the
compressed ice samples and might be of essential significance
for the formation of the amorphous matrix.
It is interesting and important to note that the pressure at which
amorphous ice can be formed is depending on the consistency
and grain--size of the initial sample state (Johari, 2000).
The lowest formation pressure of HDA is observed when compressing
the LDA matrix, i.e. when the PLS scattering gives evidence of
a strong and isotropic distribution of interfaces within the amorphous
matrix.

\newpage

Alba--Simionesco, C. and Tarjus, G. (2000).
Europhys. Lett. 52, 297--303.\\
                
Brovchenko, I. and Geiger, A. \& Oleinikova, A. (2003).
J.\ Chem\ Phys. 118, 9473.\\

Debenedetti, P.G. (2003). J.\ Phys.:\ Condens.\ Matter 15, R1669.\\

Dewhurst, C.
{\it Graphical Reduction and Analysis SANS Program},
Institut Laue Langevin, Grenoble, France, 2002.\\
                
Doremus, R.H. (1985). Rates of Phase Transitions,
Academic Press Inc. London.\\
                
Elliott, S.R. (1991). Phys.\ Rev.\ Lett. 67, 711;
(1995). J.\ Non--Cryst.\ Solids 182, 40.\\
                
Glatter, O. and Kratky, O. Small Angle X--ray Scattering, Academic Press Inc. (London) Ltd., 1982.\\
                        
Guillot, B. \& Guissani, Y. (2003). J.\ Chem.\ Phys. 119 11740.\\

Johari, G. (2000). Phys. Chem. Chem. Phys 2, 1567--1577.\\

Koza, M.M. \& Schober, H. \& Fischer, H.E. \& Hansen, T. \&
Fujara, F. (2003). J.\ Phys.:\ Condens.\ Matter 15,
321--332.\\

Koza, M.M. \& Geil, B. \& Schober, H. \& Natali, F. (2005a).
Phys.\ Chem.\ Chem.\ Phys. 7 1423--1431.\\

Koza, M.M. \& Geil, B. \& Winkel, K. \& K\"ohler, C. \&
Czeschka, F. \& Scheuermann, M. \&  Schober, H. \& Hansen, T. (2005b).
Phys.\ Rev.\ Lett. 94, 125506.\\
               
Koza, M.M. \& Hansen, T. \& May, R.P. \& Schober, H. 
(2006). J.\ Non-Cryst.\ Solids 352, 4988--4993.\\

Kurita, R. \& Tanaka, H. (2004). Science 306, 845--848.\\

Kurita, R. \& Tanaka, H. (2005). J.\ Phys.:\ Condens.\ Matter 
17, L293--L302.\\

Lindner, P. \& Zemb, T. {\it Neutron, X--Ray and Light Scattering},
North--Holland, Elsevier Science Publishers B.~V. 2002.\\

Loerting, T. \& Salzmann, C. \& Kohl, I. \& Mayer, E. \&
Hallbrucker, A. (2001). \ Phys.\ Chem.\ Chem.\ Phys. 3,
5355--5357\\

Martonak, R.\& Donadio, D. \& Parrinello, M. (2004).
Phys.\ Rev.\ Lett. 92 225702.\\

Mishima, O. (1996). Nature, 384, 546--549.\\

Mishima, O. \& Stanley, H.~E. (1998).  Nature 396, 
329--335.\\

Poole, P.H. \&  Essmann, U. \& Sciortino, F. \& Stanley, H.E.
(1993). Phys. Rev. E 48, 4605--4610.\\

Schober, H. \& Koza, M. \& T\"olle, A. \& Fujara, F. \&
Angell, C.A. \& B\"ohmer, R. (1998). Physica B 241--243,
897--902.\\

Schober, H. \& Koza, M.M. \& T\"olle, A. \& Masciovecchio, C.
\& Sette, F. \& Fujara, F. (2000). Phys.\ Rev.\ Lett. 85, 4100--4103.\\

Senker, J \& Sehnert, J. and Cornell, S. (2005).
J.\ Am.\ Chem.\ Soc. 127, 337--349.\\
                
Stanley, H.E. \& Buldyrev, S.V. \& Canpolat, M. \& 
Mishima, O. \& Sadr--Lahijanin, M.R.\& Scala, A. \& Starr, F.W. (2000).
Phys.\ Chem.\ Chem.\ Phys. 2, 1551--1558.\\

Tanaka, H. (2000). Europhys. Lett. 50, 340--346.
                
\end{document}